\documentclass[11pt,twoside]{article}

\usepackage{asp2014}

\aspSuppressVolSlug
\resetcounters

\begin{document}

\title{\texttt{LSTOSA}: Onsite processing pipeline for the CTA Larged-Sized Telescope prototype}

% Note the position of the comma between the author name and the
% affiliation number.
% Authors surnames should come after first names or initials, eg John Smith, or J. Smith.
% Author names should be separated by commas.
% The final author should be preceded by "and".
% Affiliations should not be repeated across multiple \affil commands. If several
% authors share an affiliation this should be in a single \affil which can then
% be referenced for several author names. If only one affiliation, no footnotes are needed.
% See ManuscriptInstructions.pdf and ASP's manual2010.pdf 3.1.4 for more details
\author{J.E. Ruiz$^1$, D. Morcuende$^2$, L. Saha$^2$, A. Baquero$^2$, J.L. Contreras$^2$, and I. Aguado$^2$ for the CTA LST project}
\affil{$^1$Instituto de Astrof\'{i}sica de Andaluc\'{i}a, Granada, Spain; \email{jer@iaa.es}}
\affil{$^2$IPARCOS and Department of EMFTEL, Universidad Complutense de Madrid, E-28040 Madrid, Spain}

% This section is for ADS Processing.  There must be one line per author. paperauthor has 9 arguments.
\paperauthor{J.E. Ruiz}{jer@iaa.es}{}{Instituto de Astrof\'{i}sica de Andaluc\'{i}a}{DAE}{Granada}{}{18008}{Spain}
\paperauthor{D. Morcuende}{dmorcuen@ucm.es}{}{Universidad Complutense de Madrid}{IPARCOS and EMFTEL department}{Madrid}{}{E-28040}{Spain}
\paperauthor{L. Saha}{labsaha@ucm.es}{}{Universidad Complutense de Madrid}{IPARCOS and EMFTEL department}{Madrid}{}{E-28040}{Spain}
\paperauthor{A. Baquero-Larriva}{labsaha@ucm.es}{}{Universidad Complutense de Madrid}{IPARCOS and EMFTEL department}{Madrid}{}{E-28040}{Spain}
\paperauthor{J.L. Contreras}{joseluco@ucm.es}{0000-0001-7282-2394}{Universidad Complutense de Madrid}{IPARCOS and EMFTEL department}{Madrid}{}{E-28040}{Spain}
\paperauthor{I. Aguado}{iaguado@ucm.es}{}{Universidad Complutense de Madrid}{IPARCOS and EMFTEL department}{Madrid}{}{E-28040}{Spain}

% There should be one \aindex line (commented out) for each author. These are used to
% build up the author index for the Proceedings. The surname must come first, followed by
% initials. Note the use of ~ before each initial to control spacing.
% The \author entries (see above) have surname last. These \aindex entries have
% surname first.
% The Aindex.py command will create them for you after you have constructed the \author
% The first entry should be the first author, for bold-facing the author index page-reference

%\aindex{Ruiz,~J.~E.}
%\aindex{Morcuende,~D.}
%\aindex{Saha,~L.}
%\aindex{Baquero-Larriva,~A.}
%\aindex{Contreras,~J.}
%\aindex{Aguado,~I.}

\begin{abstract}
The prototype of the Large-Sized Telescope (LST) of the Cherenkov Telescope Array (CTA) is presently going through its commissioning phase. A total of four LSTs, among others, will operate together at Observatorio del Roque de Los Muchachos, which will host the CTA North site.

A computing center endowed with 1760 cores and several petabytes disk space is installed onsite. It is used to acquire, process, and analyze the data produced, at a rate of 3~TB/hour during operation. The LST On-site Analysis \texttt{LSTOSA} is a set of scripts written in Python which connects the different steps of \texttt{lstchain}, the analysis pipeline developed for the LST. It processes the data in a semiautomatic way producing high-level data and quality plots including detailed provenance logs. Data are analyzed before the next observation night to help in the commissioning procedure and debugging.

\end{abstract}

% These lines show examples of subject index entries. At this stage these have to commented
% out, and need to be on separate lines. Eventually, they will be automatically uncommented
% and used to generate entries in the Subject Index at the end of the Proceedings volume.
% Don't leave these in! - replace them with ones relevant to your paper.
%\ssindex{pipelines}

% These lines show examples of ASCL index entries. At this stage these have to commented
% out, and need to be on separate lines. Eventually, they will be automatically uncommented
% and used to generate entries in the ASCL Index at the end of the Proceedings volume.
% The ascl.py command will scan your paper on possible code names.
% Don't leave these in! - replace them with ones relevant to your paper.

\section{Introduction}
The Cherenkov Telescope Array CTA\footnote{\url{https://www.cta-observatory.org/}} \citep{CTA}, is the next generation of ground-based Cherenkov telescopes
observing the gamma-ray sky in the energy range $20$~GeV$-300$~TeV. The array will be composed of imaging atmospheric Cherenkov telescopes of three different sizes distributed into two sites, one in the northern hemisphere in the Canary Island of La Palma (Spain) and another located in the southern hemisphere at Paranal Observatory (Chile).

The prototype for CTA of the Large-Sized Telescope LST-1 \footnote{\url{https://www.cta-observatory.org/project/technology/lst/}} \footnote{\url{http://lst1.iac.es/}} \citep{LST}, located at the \textit{Observatorio del Roque de Los Muchachos} (ORM) in La Palma, is presently going through its commissioning phase. It is placed next to the two MAGIC (Major Atmospheric Gamma Imaging Cherenkov) telescopes, which is an advantage for the operation, maintenance and calibration of the telescope. A total of four LSTs, among other different-size telescopes, will operate together at ORM as part of the CTA North (CTA-N) site as is shown in Figure \ref{cta-n}.

\articlefigure{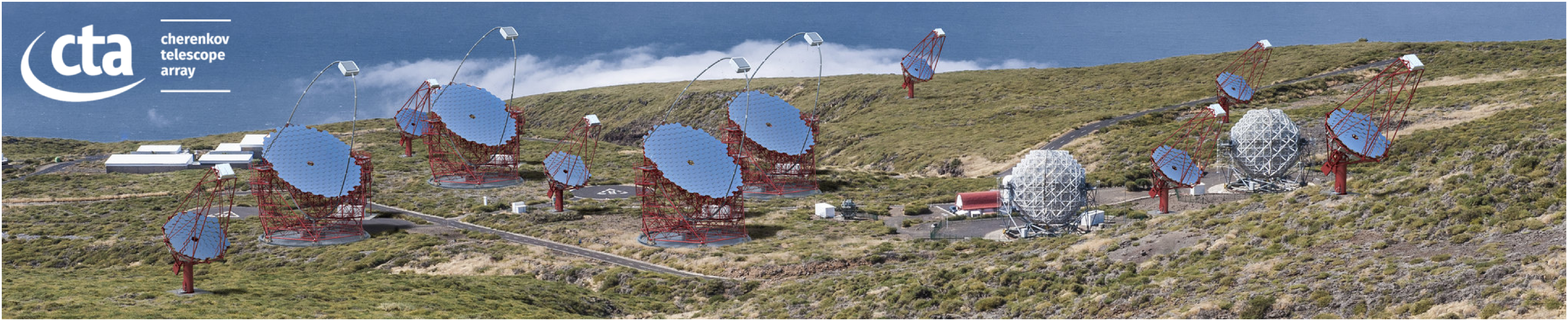}{cta-n}{Artistic rendering of the future layout of the CTA-N array at the ORM observatory at La Palma. The two existing MAGIC telescopes point towards the sea, the LST prototype stands next to them. Credit: Gabriel P\'{e}rez Diaz, IAC, SMM.}

Due to the large amount of daily recorded data, transferring the raw data online through the network connection from La Palma to continental Europe is not feasible. Therefore an LST On-Site Analysis (\texttt{LSTOSA}) pipeline is being developed to perform the reduction of the raw data at the LST site. Its aims to process all the data acquired by the telescope, providing low and intermediate level analysis products to the LST Collaboration.

\section{LST-1 data levels and characteristics}

Cherenkov telescopes record the faint and short light pulses generated in Extensive Air Showers (EAS), initiated by the interaction of primary particles (photons among them) with the atmosphere, usually called \textit{events}. The 23-m diameter LST-1 is equipped with a 1855 photomultipliers (PMT) camera providing fast sampling of the triggered light pulses. During observations the telescope records $\sim 10^4$ events per second, where the electronic behind each PMT generates 40 consecutive 1-ns samples of their \textit{waveforms} with a resolution of 12~bits. Before the observations, and also interleaved with them, a much lower rate of artificial \textit{calibration events} is also acquired.  Altogether around 3 TB of data are recorded per hour of observation time, constituting the lowest level of data, dubbed R0. The first step in the data reduction pipeline is to calibrate the R0 data, extracting the event arrival times and producing images of the light collected, which are then parametrized. Both types of information are grouped in the Data Level 1 (DL1). In the next step the physical parameters of the primary particle are inferred giving rise to DL2 data. Finally, photon event candidates are selected and the response of the telescope estimated and grouped in what is called DL3. Presently \texttt{LSTOSA} carries out the analysis from R0 to DL2, as summarized in Figure~\ref{flow_data}.

\articlefigure{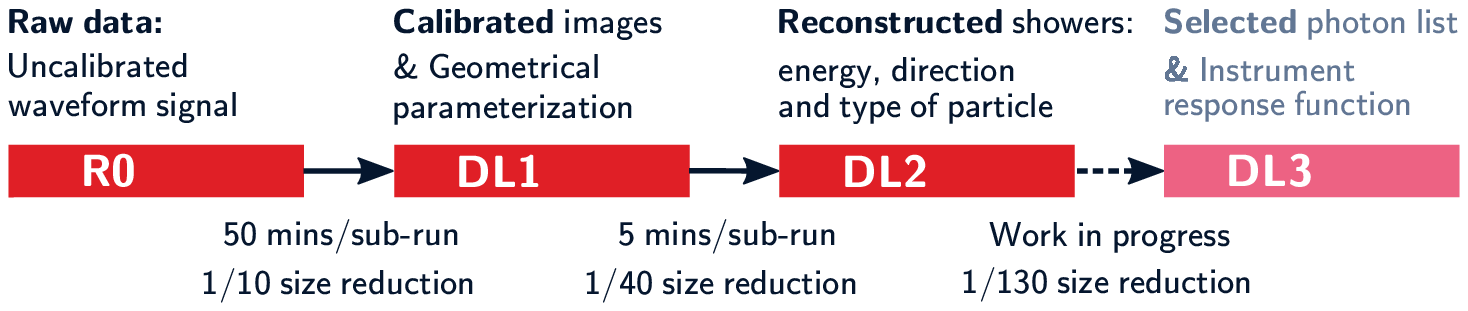}{flow_data}{Data reduction steps, starting from raw uncalibrated waveform signals to selected photon lists.}

\section{Onsite computing infrastructure}
An \textit{IT Container} housing a compact data center, placed next to the telescope, allows us to record and process the data acquired by the telescope, including \texttt{LSTOSA} pipeline data processing. The data center provides 55 computing nodes, each one with 32 cores, for a total of 1760 cores and 3.5~PB of disk space. This cluster uses the CentOS operating system, administers the work load through the SLURM batch scheduling system and implements the Fujitsu Scalable File System (\textit{FEFS}) based on Lustre.

Once the data have been recorded and processed, they are copied via the network to the computing center PIC (Port d'Informació Científica) located in Barcelona. The members of the LST Collaboration have access to the so called \textit{IT Container} and use it for the commissioning of the telescopes and preliminary astrophysics analysis. The vast computing power available in the \textit{IT Container} is key to make possible the processing of LST-1 data.

\section{LST On-Site Analysis - \texttt{LSTOSA}}

\texttt{LSTOSA} aims to run the data reduction and analysis chain up to DL3, though it currently produces calibrated data and image parameters (DL1), as well as estimates of the primary particle physical parameters (DL2). Together with the reduced data, quality check plots are provided, which help to debug potential problems and to commission the telescope. \texttt{LSTOSA} also tracks the provenance of the analysis products to help in achieving reproducibility of the results. During the science phase of the LST1 this fast offline high-level analysis will also be a tool to assist target of opportunity observations and provide science alarms.

\texttt{LSTOSA} \footnote{\url{https://gitlab.cta-observatory.org/cta-array-elements/lst/analysis/lstosa}} consists of a set of Python scripts connecting the different steps of the data reduction pipeline developed for the LST \texttt{lstchain}\footnote{\url{https://github.com/cta-observatory/cta-lstchain}} \citep{P8-84_adassxxx}. So far, it works in a semiautomatic way aiming for a fully automatized version. It is required that data acquired during an observation night should be analyzed before the next one. To achieve this goal \texttt{LSTOSA} splits the CPU-intensive data reduction steps of each observation into many jobs executed in parallel.

The workflow process starts with a summary of the observations of the night, it is then decomposed in sequences of observations and calibrations. A pilot job is built for each sequence, and they are sent to the scheduling system SLURM, which takes care of allocating the resources and provides a first level of parallelization. 

Each observation, usually called a \textit{run} is normally composed of a set of $\sim 10^{2}$ files sometimes called \textit{sub-runs}, each one comprising less than 10 seconds of data taking. The pilot jobs launch one job for each \textit{sub-run} comprising the observation, using the SLURM job array capabilities. This provides a second level of parallelization. Once all jobs are finished, the results are copied to the final storage locations and merged, where data check plots are provided to the collaboration through a web interface.

\section{Provenance}
%\begin{itemize}
%    \item \textbf{IVOA} provenance model with \textbf{W3C} syntax.
%    \item Run-wise provenance products for each data level (\texttt{.json} serialization / \texttt{.pdf} graphs).
%    \item Configuration files and input parameters kept run-wise to ensure \textbf{reproducibility}.
%    \item We plan to use a \textbf{database} and develop a provenance \textbf{inspection query tool}.
%\end{itemize}

The data analysis steps executed to create DL1 and DL2 level data are captured for each \textit{run}, together with the configuration parameters and files needed as well as intermediate files produced. This information is serialized in \texttt{.json} formatted files, following the IVOA Provenance Model Recommendation \citep{ivoa_provenance}. Provenance graphs are also provided in \texttt{.pdf} formatted files, rendering a detailed complete view of the data analysis process which improves process inspection and helps achieving reproducibility. Tracking of the calibration steps will be implemented shortly, and a more detailed provenance query tool is also foreseen, which would need to store the provenance information in a database.

\section{Conclusion}
A first version of the \texttt{LSTOSA} pipeline is currently working at the onsite processing system for the LST prototype with satisfactory results. It supports the commissioning of the telescope and the development of the analysis software on real data. It makes full use of the computing capacities of the onsite computing cluster with a high degree of automation aiming at achieving a fully automatized fast pipeline in the near future.

\acknowledgements
This work was conducted in the context of the CTA LST Project. We gratefully acknowledge financial support from the agencies and organizations listed in \url{http://www.cta-observatory.org/consortium_acknowledgments}. We also acknowledge support from Universidad Complutense de Madrid through an UCM-Harvard (CT17/17-CT18/17) grant (D.~Morcuende), Ecuadorian Institute of the Secretariat of Higher Education, Science, Technology and Innovation (SENESCYT) (A.~Baquero), and Spanish MICINN under project FPA2017-82729-C6-3-R. J. E. Ruiz acknowledges financial support from the State Agency for Research of the Spanish MCIU through the
"Center of Excellence Severo Ochoa" award to the Instituto de Astrofísica de Andalucía (SEV-2017-0709).

\bibliography{P8-138}  % For BibTex

% if we have space left, we might add a conference photograph here. Leave commented for now.
% \bookpartphoto[width=1.0\textwidth]{foobar.eps}{FooBar Photo (Photo: Any Photographer)}

\end{document}